\documentstyle[12pt,psfig]{article}
\def\reference{\parskip 0pt\par\noindent\hangindent 0.5 truecm}

\def\fm{\hbox{$.\!\!^{\rm m}$}}
\begin{document}
%
%
\title{Constraining the evolution of young radio-loud AGN}
%


\author{I.A.G. Snellen, $^{1}$ 
 K.-H. Mack, $^{2,3,6}$ 
 R.T. Schilizzi, $^{4,5}$ 
 W. Tschager $^{5}$
} 

\date{}
\maketitle

{\center
$^1$ Institute for Astronomy, University of Edinburgh, Blackford Hill
, Edinburgh EH9 3HJ, UK, ignas@roe.ac.uk\\[3mm]
$^2$ ASTRON/NFRA, Postbus 2, NL-7990 AA Dwingeloo, The Netherlands\\[3mm]
$^3$ Istituto di Radioastronomia de CNR, Via Gobetti 101, I-40129 Bologna, Italy, mack@ira.cnr.it\\[3mm]
$^4$ Joint Institute for VLBI in Europe, Postbus 2, NL-7990 AA, Dwingeloo, The Netherlands, schilizzi@jive.nl\\[3mm]
$^5$ Leiden Observatory, P.O. Box 9513, NL-2300 RA, Leiden, The Netherlands, tschager@strw.leidenuniv.nl\\[3mm]
$^6$ Radioastronomisches Institut der Universit\"at Bonn, Auf dem H\"ugel 71, D-53121 Bonn, Germany\\[3mm]
}

%
\begin{abstract}
GPS and CSS radio sources are the
objects of choice to investigate the evolution of young radio-loud 
AGN. Previous investigations, mainly based on number counts and source 
size distributions, indicate that GPS/CSS sources decrease significantly
in radio power when evolving into old, extended objects. We suggest this 
is preceded by a period of increase in radio luminosity, which lasts as 
long as the radio source is confined within the core-radius of its
host galaxy. 
We have selected a sample of nearby  compact radio sources, unbiased 
by radio spectrum, to determine their luminosity function, 
size distribution, dynamical ages, and emission line properties in a complete
and homogeneous way. 
First results indicate that the large majority of objects ($>$80\%)
exhibit classical GPS/CSS radio spectra, and show structures consistent
with them being compact double or compact symmetric objects. 
This sample provides an ideal basis to further test and constrain possible 
evolution scenarios, and to investigate the relation between 
radio spectra and morphologies, orientation and Doppler boosting 
in samples of young radio-loud AGN, in an unbiased way.
\end{abstract}


\bigskip

%
%

\section{Current views on radio source evolution}

It has been shown beyond reasonable doubt that the archetypical 
Gigahertz Peaked Spectrum (GPS) radio galaxies are young objects, 
with typical dynamical ages of $10^{2-3}$ years (Polatidis \& Conway 
2002, and references therein). These sources are therefore
 the prime candidates to be the young progenitors of classical 
large-size radio sources, which subsequently evolve into Compact 
Steep Spectrum (CSS), and FRI and/or FRII radio sources. 
Although it is not clear whether all GPS sources are young objects
(certainly the GPS quasars seem to be an unrelated distinct 
 class of objects), GPS galaxies are the key objects to 
study the early evolution of radio-loud AGN. 

Current views on radio source evolution are mainly based on 
the radio power - linear size ($P-D$) diagram, and source size
distributions. These suggest that if GPS and CSS sources evolve
into large size radio sources, they should decrease in radio power
by at least an order of magnitude, to account for their 
relatively high space density 
(Fanti et al. 1995; Readhead et al. 1996, O'Dea \& Baum 1997).  
This luminosity evolution scenario is backed up by theory, ie. it is 
expected that a radio source with a constant jet power, expanding in
 a medium with a declining density profile, will decrease 
in radio power (eg. Baldwin 1982; 
Kaiser \& Alexander 1997). 

\subsection{Perspective from our group}

We have used the existing samples of Stanghellini et al. (1998; bright GPS galaxies), Fanti et al. 1990 (bright CSS galaxies), and 
Snellen et al. (1998a; faint GPS galaxies) to study early radio source
evolution. This has been described in detail in Snellen et al. 
(2000). We noticed a distinct difference between the redshift distribution
of GPS galaxies from the Stanghellini et al. sample and 
3C galaxies, with the GPS galaxies biased towards higher
redshifts (95\% confidence in the KS test). First of all, this makes
the number count analyses mentioned above, which are averaged over
a large redshift range, less straightforward. Furthermore, it is 
important to understand how two populations, which evolve into 
each other on timescales several orders of magnitude less than the 
Hubble time, can have different redshift distributions. 
It is clear that to explain this redshift bias in a flux density 
limited sample, the luminosity function of GPS galaxies has to be
different from that of old, extended objects, if the former are to 
develop into the latter.

\begin{figure}
 \centerline{
 \psfig{file=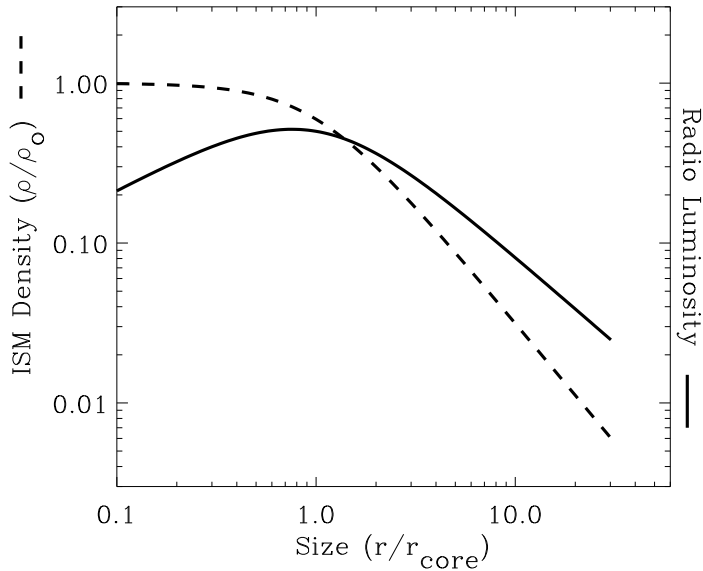,width=10cm}}
 \caption{Typical density profile of a galaxy (dashed line), and 
the expected luminosity evolution as function of size for a radio source (solid line), according to Snellen et al. (2000).}
 \label{fig1}            
 \end{figure}

We advocate that it is the qualitatively different
individual luminosity development of young radio sources compared to 
that of old objects which causes a flatter overall luminosity 
function. If radio sources increase in radio power until
they grow beyond a certain size and thereafter decrease in radio 
power, a randomly aged population of compact radio sources will be
biased towards higher luminosities than a population of large-size 
objects, producing a flatter luminosity function. This behaviour 
in luminosity evolution is expected from theory, in which the 
density profile of the core of the host galaxy is flat (see Fig. \ref{fig1}, 
and 
the density only starts to decrease outside this core radius
(Snellen et al. 2000; Alexander 2000; O'Dea 2002).
Some new evidence for this scenario is also presented 
by Perucho \& Marti 2002a,b).

\subsection{Synchrotron self absorption and self-similar evolution}

Snellen et al. (2000) used the three formerly mentioned 
samples to study the relation between the radio spectra and 
the overall sizes of the GPS and CSS sources.
When assuming Synchrotron self absorption (SSA), the 
frequency and flux density of the overall spectral turnover 
provide a good indication of the subtended solid angle of the radio 
source and subsequently of the size of the dominant radio components,
with the angular size, $\theta_{\rm{ssa}}$, being, 
\[ \theta_{\rm{ssa}}\propto B^{1/4}S^{1/2}_{\rm{peak}}\nu^{-5/4}_{\rm{peak}}\]
where $B$ is the magnetic field strength. If the weak dependence 
on $B$ is taken out by assuming an equipartition magnetic field, 
then the SSA components sizes are found to be linearly proportional
to the {\it overall} size of the GPS/CSS sources as measured from 
VLBI, over a luminosity and linear size range of several orders of 
magnitude. This indicates that the ratio of overall-size to 
lobe-size stays constant while the sources grow, and that therefore 
young radio sources evolve in a self-similar way. Furthermore, it 
provides strong
evidence that the turnovers in the radio spectra of GPS/CSS sources
are indeed caused by SSA, and not free-free absorption (FFA). 
Further evidence in favour of this scheme is presented by
Tschager et al. (2002), who find that a new sample of very
faint CSS galaxies also follow the relation given above. 

Although it is shown that FFA does occur in some compact GPS sources 
(eg. Kameno et al. 2002; Inoue et al. 2002; Vermeulen et al., 2002),
it seems that even in these objects FFA only generally affects faint 
components close to the nucleus, and not the mini-lobes which form 
the dominant components in the overall radio spectrum.
Although it is possible to fit the spectra of individual sources 
with fine-tuned FFA models, it is not clear to us why the 
conditions of FFA 
would change in such way over a large range of radio source
luminosities and linear sizes, that they exactly mimic the 
frequency and strength of the turnovers as expected for SSA. 
This makes a very strong case for SSA being the dominant absorption 
process in GPS and CSS galaxies. Highly accurate flux density 
monitoring can now be used to further test self-similar 
evolution by comparing the long-term flux density 
variability with what is expected for this scenario 
(eg. Tingay et al. 2002).

\subsection{Open questions}

Although we now start to have a more or less general idea of  
how GPS and CSS could evolve into old, extended objects, many
questions remain open, and the proposed evolution scenarios should 
be further tested. Is the initial phase of increase in luminosity 
correct, and if so, how strong is this rise? 
Do all GPS/CSS galaxies evolve into extended objects, 
or is there a significant population of short-lived objects, eg.~due 
to the lack of fuel (these may show up as dying drop-outs in 
GPS/CSS samples; eg. Marecki et al., 2002)?
Does intermittance play an important role? Do GPS and CSS
sources evolve into FR II only, or do they also evolve into in FR Is
as well?  
Can we in some way recognise objects which are bound to evolve 
in either class of object, or are FR I and II physically so closely
linked (eg. through evolution) that there is no difference in their
progenitors? 

Several other matters have to be settled. The link between
the radio spectra and VLBI morphologies have to be further 
investigated. Do all GPS galaxies have compact symmetric (CSO)
morphologies, such as the bright archetypes? What effect could 
orientation have on the classification of an object as either a 
GPS source or a CSO (eg. Snellen et al. 1998b)? Is Doppler boosting
really
not important in these objects, as is always assumed, or does 
it significantly affect our statistical analyses? And even more 
general, but not too obvious, how do you exactly decide what is and what is 
not a GPS source? To address many of these issues, we are studying a new 
sample of nearby, young radio-loud AGN.
 

\section{A new, low-redshift sample of young radio sources}
\subsection{Rationale}

For a statistically sound analysis of radio-loud AGN, it is vital
to study a complete, volume limited sample, unaffected by possible 
cosmological evolutionary effects. Evidently, a {\it nearby} sample
of young radio sources would be the obvious choice, since
key observations, such as optical spectroscopy, 
are much easier to obtain, and the statistics and properties of other 
(older) classes of radio sources are well established at low redshift.
Radio source evolution scenarios as proposed above provide distinct 
predictions on luminosity functions, linear size distributions, 
and emission line properties and dynamical ages as function of 
radio power and size. The determination of these properties will 
therefore be the main goal of our investigation, which will form 
a powerful tool to constrain evolution models.

In addition, as a spin-off, a complete sample of nearby young 
radio-loud AGN allows us to undertake unique studies of their 
HI absorption distribution (using VLBI in L-band), X-ray, infrared
and optical cluster properties. This will undoubtedly give new insights
in the intra- and intergalactic environments of young radio-loud AGN.

\subsection{Sample selection}

 \begin{figure}
\centerline{
 \psfig{file=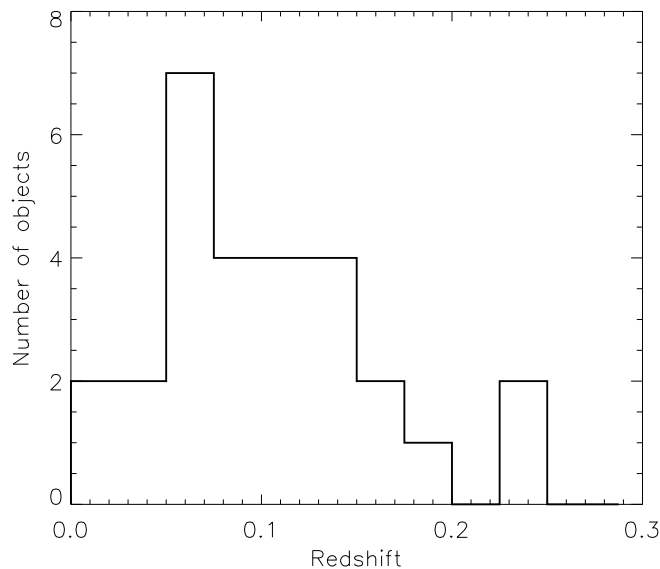,width=10cm}}
 \caption{The redshift distribution of the sources in the new sample
of nearby young radio-loud AGN.}
 \label{fig2}            
 \end{figure}

Selection of a comprehensive nearby sample has only become possible 
since the construction of the new generation of radio surveys. 
Especially the 1.4-GHz FIRST survey (White et al. 1997) with its
high resolution is ideal for 
selecting compact radio sources, and therefore forms the basis of our 
sample. Only the region north of declination +30$^\circ$ was used, 
so it overlaps with the WENSS at 325 MHz 
(Rengelink et al. 1997). 

The first step was the selection of all sources in FIRST with an 
angular size of $<2''$ and a flux density $S_{\rm{1.4GHz}}>100$ mJy. 
It is important to note that {\it no selection was made on radio 
spectra, but only on angular size}, allowing in a later stage an 
unbiased investigation of possible orientation and/or beaming effects 
on the radio spectra and morphologies of young radio-loud AGN. 
Unfortunately, by selecting at 1.4 GHz, sources with spectral 
turnovers at higher frequencies (very compact objects) may have been
missed. To catch these, sources from the flat spectrum CLASS survey 
(with S$_{\rm{5GHz}}$$>$30 mJy; Myers et al. 1995) were added to the 
sample, with an 
8.4-5 GHz spectral index such that their 1.4-GHz flux density would 
have been brighter  than 100 mJy, if they were not synchrotron self 
absorbed.

The positions of the remaining objects were cross-correlated with 
the optical APM/ POSS-I catalogue, and only those objects which coincided
within 10$''$ with a galaxy with a red magnitude $<$16\fm5 were selected. 
This yielded a total of 55 objects, which were subsequently correlated 
with other surveys such as 4C, NVSS (Condon et al. 1998), GB6 
(Gregory et al. 1996),
and CLASS, and images were obtained from the NVSS and the Digitized 
Sky Survey (DSS, Lasker et al. 1990) to check whether the optical ID 
was genuine, and to see whether any large scale radio emission may have 
been missed (on which basis they would be rejected). The literature was 
searched for any available redshifts, 
and optical spectra were taken for the remaining objects with the 
Calar-Alto 2.2-m and the 2.5-m Isaac Newton telescopes.

The final sample contains 28 compact radio sources in the 
redshift range $0.008<0.232$. The redshift distribution is shown in 
Figure \ref{fig2}.

\subsection{First observations and results}

 \begin{figure}
 \centerline{
 \psfig{file=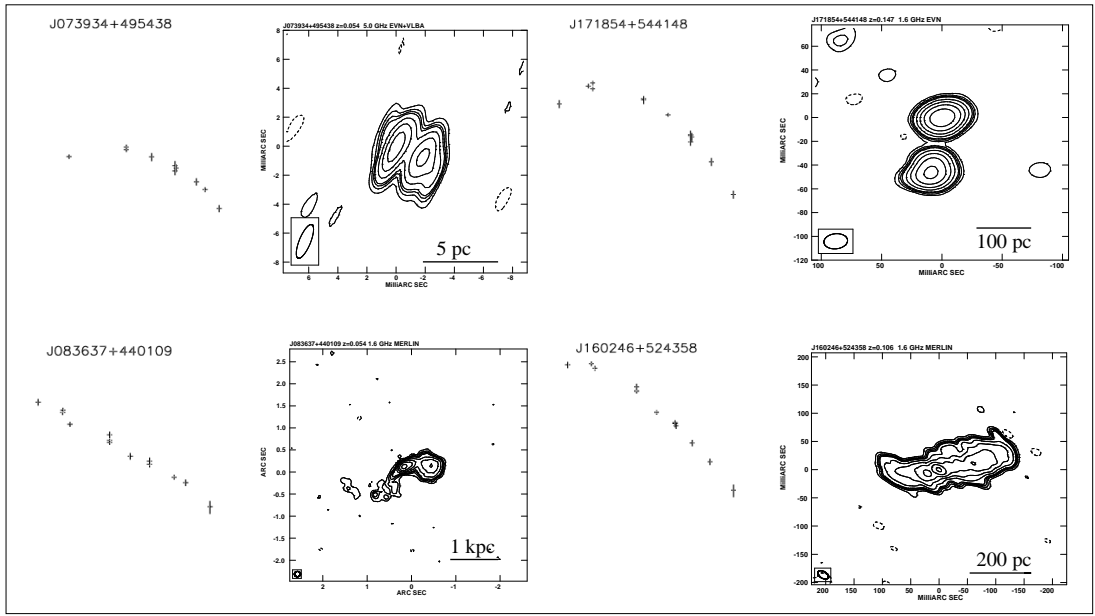,width=17cm,angle=-90}}
 \vspace{-1cm}
 \caption{Example radio spectra and morphologies of sources in 
the new sample of nearby young radio-loud AGN.}
 \label{fig3}            
 \end{figure}

A comprehensive observational programme is underway to study the
individual objects in detail. VLA observations at 5, 8.4 and 15 GHz, and 
Effelsberg observations at 2.7, 4.8, 10.5, and 32 GHz were taken to 
refine the radio spectra and to determine possible flux density 
variability. The shapes of the radio spectra have given us 
a first indication of the angular sizes expected for these objects
(see section 1.2). This information was used to divide the sample into
small-size, intermediate-size, and large-size sub-samples. 
These sub-samples were then
observed with the VLBA+EVN at 5 GHz ($\sim$1.5 mas resolution), the 
EVN at 1.6 GHz ($\sim$15 mas resolution), and MERLIN at 1.6 GHz 
($\sim$150 mas resolution) respectively. 
So far, all but one source have been observed in 
this way. In addition, optical imaging in B, V, R, and I band filter,
and HI absorption observations with WSRT are underway. 

The first results look very promising. More than 80\% of the objects
have radio spectra which can be classified as classical 
GPS or CSS sources (see Fig. \ref{fig3}), 
with the remainder also showing a spectral peak
but clearly exhibiting variability. Most of the objects have 
 morphologies consistent with them being compact double or compact 
symmetric objects, as expected for this class of objects, 
but some seem to show core-jet or more complex structures.
This sample is ideal to further study the relation between radio 
spectrum, morphology, orientation, and Doppler boosting.

At this stage, the foundation has been laid to determine directly 
the local luminosity function, source size distribution, and the 
dynamical age distribution for young radio-loud AGN, without
 cosmological bias.


%
%





\section*{Acknowledgements}

We wish to thank Tasso Tzioumis for the excellent organisation of 
the workshop, and the villagers of Kerastari for their warm
hospitality. This paper is partly based on observations collected at the 
German-Spanish Astronomical Center, Calar Alto, Spain, 
operated by the Max-Planck-Institute f\"{u}r Astronomie, Heidelberg, 
jointly with the Spanish National Commission for Astronomy; the Isaac Newton 
Telescope (through the service programme), which is
operated on the island of La Palma by the Isaac Newton Group in the Spanish
Observatorio del Roque de los Muchachos of the Instituto de Astrofisica de 
Canarias; The European VLBI Network (EVN, making use of 
EC's Access to Research Infrastructures Programme,
under Contract No. HPRI-CT-1999-00045), which is a joint facility of European,
Chinese and other radio astronomy institutes funded by their national research 
councils; MERLIN, a National Facility operated by the University of Manchester at Jodrell Bank Observatory on behalf of PPARC. The National Radio Astronomy Observatory is
a facility of the National Science Foundation operated under cooperative agreement by Associated Universities, Inc. 
KHM was supported by a Marie-Curie fellowship of the European Commission.


\section*{References}






\reference Alexander P. 2000, MNRAS, 319, 8
\reference Baldwin J., 1982, In: Extragalactic radio sources,
           NM, D. Reidel Publishing Co., 1982, p. 21
\reference Condon J.J., Cotton W.D., Greisen E.W., Yin Q.F., Perley R.A.,
           Taylor G.B., Broderick J.J., 1998, Astronomical Journal, 
           {\bf 115},  1693
\reference Fanti R., Fanti C., Schilizzi R.T., Spencer R.E., Nan Rendong, 
           Parma P., Van Breugel W.J.M., Venturi T., 1990, A\&A, 231, 333
\reference Fanti C., Fanti R., Dallacasa D., Schilizzi R.T., Spencer R.E.,
           Stanghellini C., 1995, A\&A, 302, 317
\reference Gregory P.C., Scott W.K., Douglas K., Condon J.J., 1996,
           Astrophys. J. Suppl. {\bf 103}, 427
\reference Inoue et al. 2002, PASA, this volume
\reference Kaiser C.R., Alexander P., 1997, MNRAS, 286, 215
\reference Kameno et al. 2002, PASA, this volume
\reference Lasker, B.M., Sturch C.R., McLean B.J., Russell, J.L., Jenkner H.,
           Shara M.M., 1990, Astronomical Journal, {\bf 99}, 2019
\reference Marecki et al. 2002, PASA, this volume
\reference Myers S.T., Fassnacht C.D., Djorgovski S.G., Blandford R.D.,
             Matthews K., Neugebauer G., Pearson T.J., Readhead A.C.S., 
             Smith J.D., Thompson D.J., Womble D.S., Browne I.W.A., 
             Wilkinson P.N., Nair S., Jackson N., Snellen I.A.G., Miley G.K., 
             de Bruyn A.G., Schilizzi R.T, 1995,
             {\it Astrophysical Journal}, {\bf 447}, L5
\reference O'Dea C.P., Baum S.A., 1997, AJ, 113, 148
\reference O'Dea et al. 2002, PASA, this volume
\reference Perucho M, Marti J.M., 2002a, ApJ 568, 639
\reference Perucho M, Marti J.M., 2002b, PASA, this volume
\reference Polatidis A. \and Conway J. 2002, PASA, this volume
\reference Readhead A.C.S., Taylor G.B., Xu W., 
           Pearson T.J., Wilkinson P.N., 1996, ApJ, 460, 634
\reference Rengelink R.B., Tang Y., de Bruyn A.G., Miley G.K., Bremer M.N., 
         R\"ottgering H.J.A., Bremer M.A.R., 1997, 
         Astr. \& Astrophys. Suppl. {\bf 124}, 259
\reference Snellen I.A.G., Schilizzi R.T., de Bruyn A.G., Miley G.K.,
           Rengelink R.B., R\"ottgering H.J.A., Bremer, M.N., 1998a,
           A\&AS, 131, 435
\reference Snellen I.A.G., Schilizzi R.T., de Bruyn A.G., Miley G.K., 1998b,
           A\&A 333, 70
\reference Snellen, I. A. G.; Schilizzi, R. T.; Miley, G. K.; de Bruyn, 
           A. G.; Bremer, M. N.; R\"ottgering, H. J. A., 2000, MNRAS, 319, 445
\reference Stanghellini C., O'Dea, C. P., Dallacasa, D., Baum, S.A., 
           Fanti, R., Fanti, C., 1998, A\&AS, 131, 303
\reference Tingay S. et al. 2002, PASA, this volume
\reference Tschager et al. 2002, PASA, this volume
\reference Vermeulen et al. 2002, PASA, this volume
\reference White R.L., Becker R.H., Helfand D.J., Gregg M.D., 1997, 
           Astrophysical Journal {\bf 475}, 479


\end{document}